\documentclass[useAMS,usenatbib,usegraphicx]{mn2e}

\title[Molecular Gas in Intermediate Redshift ULIRGs]{Molecular Gas in Intermediate Redshift ULIRGs}
\author[R. Braun et al.]{R. Braun$^{1}$\thanks{E-mail:
Robert.Braun@csiro.au}, A. Popping$^{1,2}$ K. Brooks$^{1}$ and F. Combes$^{3}$\\
$^{1}$CSIRO Astronomy and Space Science, PO Box 76, Epping, NSW 1710, Australia\\
$^{2}$ICRAR, The University of Western Australia, Crawley, WA 6009, Australia \\
$^{3}$Observatoire de Paris, 61 Av. de l’Observatoire, 75014 Paris, France}
\begin{document}

\date{Accepted  Received ; in original form }

\pagerange{\pageref{firstpage}--\pageref{lastpage}} \pubyear{2011}

\maketitle

\label{firstpage}

\begin{abstract}
We report on the results of observations in the CO(1--0) transition of
a complete sample of Southern, intermediate redshift (z = 0.2 -- 0.5)
Ultra-Luminous Infra-Red Galaxies using the Mopra 22m telescope. The
eleven ULIRGs with $L_{FIR} > 10^{12.5}$ L$_\odot$ south of $\delta =
-12^\circ$ were observed with integration times that varied between 5
and 24 hours. Four marginal detections were obtained for individual
targets in the sample. The ``stacked'' spectrum of the entire sample
yields a high significance, 10$\sigma$ detection of the CO(1--0)
transition at an average redshift of z~=~0.38. The tightest
correlation of $L_{FIR}$ and $L_{CO}$ for published low redshift ULIRG
samples ($z <$ 0.2) is obtained after normalisation of both these
measures to a fixed dust temperature. With this normalisation the
relationship is linear. The distribution of dust-to-molecular hydrogen
gas mass displays a systematic increase in dust-to-gas mass with
galaxy luminosity for low redshift samples but this ratio declines
dramatically for intermediate redshift ULIRGs down to values
comparable to that of the Small Magellanic Cloud. The upper envelope to
the distribution of ULIRG molecular mass as function of look-back time
demonstrates a dramatic rise by almost an order of magnitude from the
current epoch out to 5 Gyr. This increase in maximum ULIRG gas mass with
look-back time is even more rapid than that of the star formation rate
density.
\end{abstract}

\begin{keywords}
Galaxies: high redshift -— Galaxies: ISM -— Galaxies: starburst -— Radio lines: Galaxies
\end{keywords}

\section{Introduction}
\label{sec:intro}
The Universe has evolved dramatically with cosmic time. The star
formation rate density, in particular, first rose in the first few Gyr
and then declined by more than one order of magnitude in the past
10~Gyr \citep[eg.][]{hopk06}. At the same time, the galaxies hosting
star formation have systematically changed from being dominated by the
highest mass systems in the distant past to a more equal mix of galaxy
masses \citep{heav04} in the most recent 0.5 Gyr, where a
representative local volume has been sampled. The processes driving
this dramatic evolution will only be understood if there is a more
complete documentation of the time evolution of the major baryonic
constituents of the cosmos. 

An important baryonic constituent to consider in the context of
galaxies is the molecular gas mass; since this is a prerequisite for
star formation and as such plays a key role in determining its rate
and location. While ideally Nature would have provided a direct tracer
of molecular hydrogen mass under all physical conditions, this is
unfortunately not the case; making it necessary to infer molecular gas
mass from indirect measures. An important proxy for molecular hydrogen
is the detection of carbon monoxide (CO) emission lines. Despite the
fact that the CO emission lines are often highly self-opaque, high
resolution studies of individual molecular clouds and nearby external
galaxies have permitted at least rough calibration of the total
molecular hydrogen mass that is statistically associated with an
observed CO line luminosity \citep[eg.][]{dick86}. The situation is
complicated by a significant statistical scatter as well as the likely
dependencies of this association on many other factors. A good recent
overview of such calibration factors for the CO(1--0) transition is
given in \citet{tacc08}.

The relatively high brightness of the CO(1--0) transition has
permitted its routine detection in emission at z $\sim$ 0.1
\citep[eg.][]{chun09} in a few cases out to redshifts as large as
z~$\sim$~0.4 \citep{geac09,geac11}, and most recently in very luminous
galaxies at significantly higher redshifts as well
\citep{riec10,ivis11}. Higher $J$ transitions of CO have been detected
out to redshifts of 1 $<$ z $<$ 3.5 \citep[eg.][]{genz10}. In this
paper we address a major gap in our observational knowledge of the
molecular gas mass associated with galaxies by targeting a complete
Southern sample of the eleven most luminous FIR detected galaxies at
redshifts 0.2 $<$ z $<$ 0.5 for deep integrations with the Mopra 22m
telescope. Only a handful of detections have yet been made in this
redshift range \citep{solo97,geac09,geac11}. Our study complements a
similar program undertaken by \citet{comb11} directed at 30 Northern
and equatorial targets. Together these studies provide some insights
into the evolution of the molecular gas mass in ULIRGs during this
pivotal era in cosmic history.

The current paper is organised as follows. We begin with a brief
definition of the sample in \S\ref{sec:sample}, describe the
observations and data reduction methods in \S\ref{sec:obs} and present
the results in \S\ref{sec:results}. A more extensive discussion of the
results in a cosmological context is deferred to a subsequent
publication. In this paper we adopt a flat cosmological model with
$\Omega_\lambda$~=~0.73 and a Hubble constant of
71~km~s$^{-1}$Mpc$^{-1}$ \citep{hins09}.

\section{Sample definition}
\label{sec:sample}
Studies of molecular gas in the local Universe \citep[eg.][]{solo97}
have demonstrated that the largest gaseous reservoirs are apparently
associated with the galaxies most luminous in the Far Infrared
(FIR). This correlation provides an opportunity to sample the
molecular content at higher look-back times by targeting
Ultra-Luminous Infra-Red Galaxies (ULIRGs, defined by log(L$_{FIR}$)
$>$ 12) in the relevant redshift interval for deep
integrations. Estimates of the total molecular gas mass can then be
made on the basis of the molecular mass function \citep[][]{kere03},
which may evolve, and the space density of the observed target
population. While this approach is no substitute for complete sampling
of the molecular mass function at each look-back time, it provides a
first indication of possible trends.

Our sample was defined by considering all galaxies tabulated
within the NASA/IPAC Extragalactic Database (NED)\footnote{The
  NASA/IPAC Extragalactic Database (NED) is operated by the Jet
  Propulsion Laboratory, California Institute of Technology, under
  contract with the National Aeronautics and Space Administration.}
with FIR detections at 60 and 100 $\mu$m and spectroscopic
redshifts in the range z = 0.2 -- 0.5 that exceeded a limiting FIR
luminosity log(L$_{FIR}$) $>$ 12.5. The FIR luminosity has been defined
by L$_{FIR}$ = 4 $\pi$D$_L^2$ CC F$_{FIR}$, in terms of the luminosity
distance D$_L$, a Color Correction factor, CC~=~1.42 \citep{sand96}
and a FIR flux density, F$_{FIR}$ = 1.26$\times10^{-14}$(2.58
S$_{60}$+S$_{100}$)~W~m$^{-2}$ \citep{sand96}. The eleven targets below a
Declination, $\delta < - 12^\circ$, that satisfied these requirements
are listed in Table~\ref{tab:sample}. The Declination cut-off of the
sample was chosen to be complementary to the Northern sample of
\citet{comb11}.

\begin{table}
 \centering
  \caption{The southern ULIRG sample.}
\label{tab:sample}
  \begin{tabular}{@{}lccc@{}}
  \hline
 Name       & RA$_{2000}$ & Dec$_{2000}$ & z \\
 \hline
F00320-3307 & 00:34:28.5 & $-$32:51:13.0 & 0.439   \\
00397-1312 &  00:42:15.5 & $-$12:56:03.0 & 0.26172 \\
00406-3127 &  00:43:03.2 & $-$31:10:49.0 & 0.3424  \\
02262-4110 &  02:28:15.2 & $-$40:57:16.0 & 0.49337 \\
02456-2220 &  02:47:51.3 & $-$22:07:38.0 & 0.296   \\
03538-6432 &  03:54:25.2 & $-$64:23:45.0 & 0.3007  \\
F04565-2615 & 04:58:34.7 & $-$26:11:14.0 & 0.490   \\
07380-2342 &  07:40:09.8 & $-$23:49:58.0 & 0.292   \\
23515-2917 &  23:54:06.5 & $-$29:01:00.0 & 0.3349  \\
F23529-2119 & 23:55:33.0 & $-$21:03:09.0 & 0.42856 \\
F23555-3436 & 23:58:06.5 & $-$34:19:47.0 & 0.490   \\
\hline
\end{tabular}
\end{table}

\begin{table*}
 \centering
 \begin{minipage}{170mm}
  \caption{Source attributes and results.}
\label{tab:source}
  \begin{tabular}{@{}lccccrcrrrr@{}}
  \hline
 Name       & z & L$_{FIR}$ & T$_D$ & M$_D$ & $\tau$  & $\Delta T_A^*$  & S$_{CO}$  & $L_{CO}$ & L'$_{CO}$ & M$_{H2}$\\
\ & \ & (log(L$_{\sun}$)) & (K) & (log(M$_{\sun}$)) & (min) & (mK)$^\dagger$ & (Jy-km/s)$^\ddagger$ &
(log(L$_{\sun}$)) & (log(K-\ \ \  & (log(M$_{\sun}$))\\
\ & \ & \ & \ & \ & \ & \ & \ & \ & km/s pc$^2$)) & \\
 \hline
F00320-3307 & 0.439   & 12.68 & 46 & 8.29 & 340 & 0.63   & $<$7.9 & $<$6.58 &$<$10.89 & $<$11.10\\
00397-1312 &  0.26172 & 12.67 & 52 & 8.02 & 285 & 0.61 & 12.6$\pm$4.1 & 6.32: & 10.63: & 10.79:\\
00406-3127 &  0.3424  & 12.58 & 50 & 8.03 & 283 & 0.69  & $<$7.7 & $<$6.35 & $<$10.66 &$<$10.84\\
02262-4110 &  0.49337 & 12.69 & 56 & 7.82 & 667 & 0.60 & $<$7.8 & $<$6.69 & $<$11.00 &$<$11.13\\
02456-2220 &  0.296   & 12.50 & 46 & 8.17 & 560 & 0.51   & $<$5.8 & $<$6.10 & $<$10.41 &$<$10.63\\
03538-6432 &  0.3007  & 12.58 & 49 & 8.07 & 377 & 0.50  & 11.1$\pm$3.0 & 6.39: & 10.70: &10.89:\\
F04565-2615 & 0.490   & 12.62 & 49 & 8.11 & 915 & 0.64   & 7.8$\pm$3.1 & 6.68: & 10.99: &11.18:\\
07380-2342 &  0.292   & 12.78 & 37 & 9.05 &1485 & 0.40   & $<$3.8 & $<$5.90 & $<$10.21 &$<$10.52\\
23515-2917 &  0.3349  & 12.54 & 46 & 8.17 & 455 & 0.49  & 8.5$\pm$2.8 & 6.37: & 10.68: &10.89:\\
F23529-2119 & 0.42856 & 12.52 & 47 & 8.11 & 440 & 0.48 & $<$4.1 & $<$6.28 & $<$10.59 &$<$10.80\\
F23555-3436 & 0.490   & 12.67 & 46 & 8.31 & 552 & 0.65   & $<$7.5 & $<$6.66 & $<$10.97 & $<$11.18\\
Average     & 0.38$\pm$0.09   & 12.63 & 47 & 8.29 &  & & 4.84$\pm$0.48 & 6.45 & 10.76 & 10.96\\
\hline
\end{tabular}
$^\dagger$The spectral RMS is listed for a resolution of about 90 km/s.\\
$^\ddagger$Upper limits are 2$\sigma$ for a 500 km/s linewidth.
\end{minipage}
\end{table*}

\section{Observations and Reduction}
\label{sec:obs}

Observations were carried out on the Mopra 22m telescope\footnote{The
  Mopra radio telescope is part of the Australia Telescope which is
  funded by the Commonwealth of Australia for operation as a National
  Facility managed by CSIRO.} near Coonabarabran, Australia, between
15 May and 29 October 2008. Centre frequencies for the target galaxies
varied between 77 and 91 GHz. The centre frequencies were placed in
the middle of one of four intermediate frequency bands, each of 2 GHz
width, and each sampled by 8192 spectral channels in two perpendicular
polarisations. System temperatures were measured continuously by
calibration against a noise diode and varied between about 160 and
350~K during useful observing conditions. Calibration of the antenna
temperature for atmospheric attenuation was updated at 15 minute
intervals by the use of an absorbing paddle at ambient temperature
placed over the feed \citep[employing the method of][]{ulic76}. The
telescope pointing was updated each hour using the cataloged SiO maser
of smallest angular separation with the target galaxy, yielding a
pointing accuracy of 5 -- 10 arcsec, relative to a FWHM beamwidth of
about 36 arcsec. 

Standard target observing sessions lasted one hour
and consisted of alternate target and reference spectra, each of one
minute duration, using a pair of reference locations offset by 15
arcmin to both the East and West of each target galaxy. Each target spectrum
was calibrated with the average of the two adjoining offset reference
spectra. Only those observing sessions with stable system temperatures
below about 350~K were retained for further processing.

Individual one minute integrations on the target were subjected to
Fourier filtering. Peaks in the Fourier spectrum exceeding 10 times
its RMS fluctuation level were replaced with zero. A more rigorous
Fourier clipping was then applied to narrow features in the Fourier
spectrum. Localised Fourier peaks, exceeding the average Fourier
amplitude over a sliding ten pixel window by 5 times the RMS
fluctuation level were clipped at this 5$\sigma$ excursion level. Tests
of this procedure verified that injected signals of the anticipated
amplitude and linewidth of plausible detections of our target galaxies
were not significantly degraded by this filtering. The typical
improvement in the RMS fluctuation level provided by this filtering
was 50\%. The target integrations of individual observing sessions
were averaged and a second order polynomial baseline was fit to the
central 1.75 GHz of the band and subtracted from the entire spectrum.
All observing sessions for individual targets were averaged together
with an inverse variance weighting. No further baseline fitting of any
type was applied to the averaged spectrum. 

Total observing times per
target varied between 5 and 24 hours. Spectral smoothing of the final
combined spectra provides a decreasing RMS fluctuation level that
scales approximately with the square root of bandwidth through about
30~km/s. Further spectral smoothing, from 30 to 90~km/s decreases the
fluctuation level by only about 50\%, rather than the expected 70\%,
implying that residual systematic bandpass effects begin to
dominate. The resulting RMS sensitivity at a spectral resolution of
about 90~km/s varied between 0.4 and 0.7 mK in terms of calibrated
antenna temperature, $T^*_A$. Main beam brightness temperature is
given by $T_B = T^*_A/\eta_{MB}$ and the main beam efficiency at 77 --
91 GHz is estimated to be $\eta_{MB}~\sim~0.49$ \citep{ladd05}. The
assumed telescope gain that relates calibrated antenna temperature to
flux density in this frequency range\footnote{Calibration factors are
  documented in the Mopra guide
  http://www.narrabri.atnf.csiro.au/mopra/mopragu.pdf} is 22~Jy/K.

\begin{figure*}
\begin{minipage}{170mm}
\includegraphics[width=170mm]{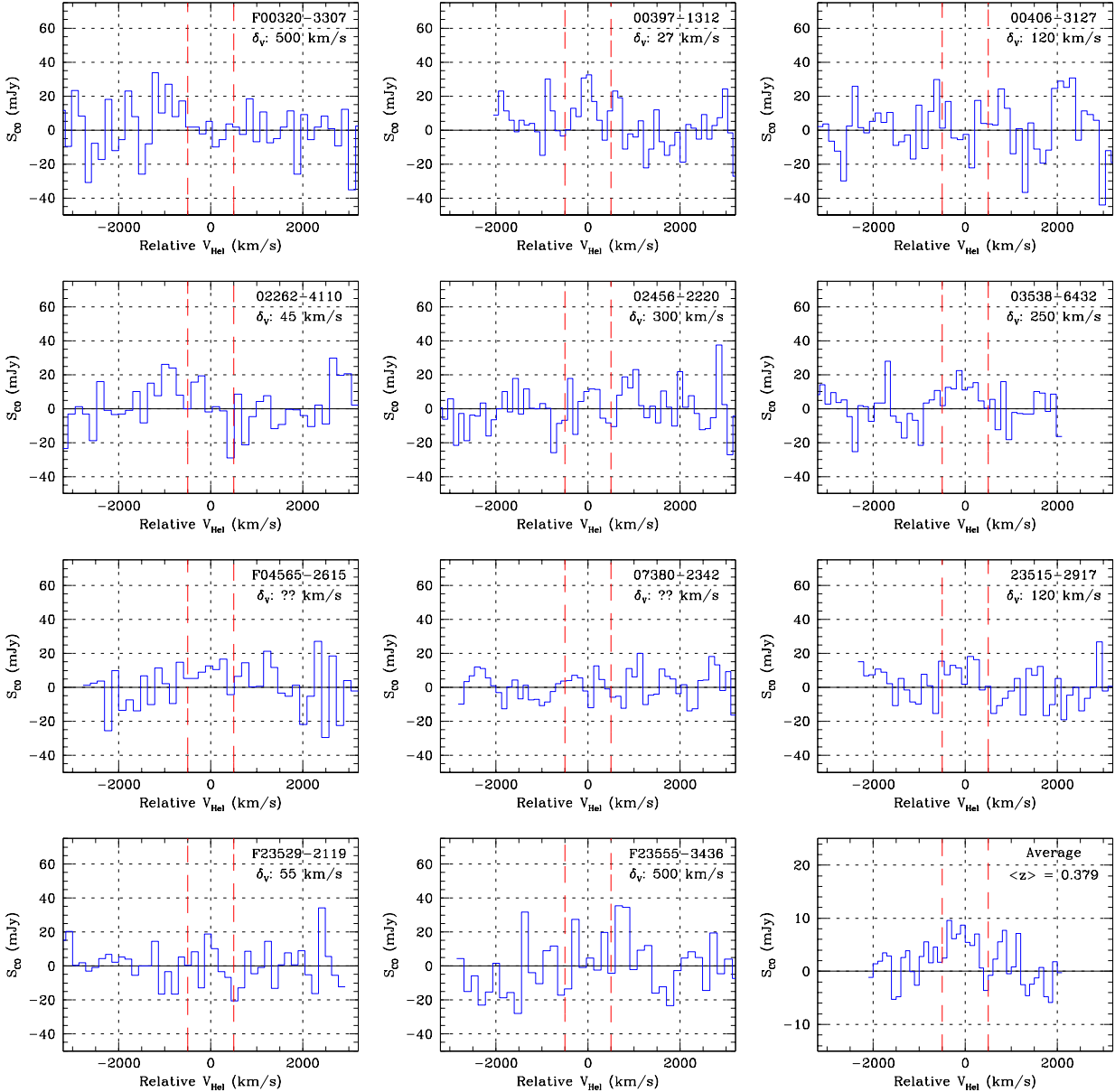}
\caption{Final spectra of CO(1--0) for the target galaxies in the
  sample shown at a spectral resolution of about 90 km/s. Vertical lines mark
  an interval of 1000 km/s centred on the expected systemic
  velocity. The published uncertainty in the systemic velocity is
  indicated in each panel. The average stacked spectrum is shown at
  lower right with the average redshift indicated.}
\label{fig:spectra}
\end{minipage}
\end{figure*}

\section{Results}
\label{sec:results}

Final spectra for each of the observed targets are shown in
Fig.~\ref{fig:spectra} at a spectral resolution of about
90~km/s. Vertical lines in this figure mark an interval of 1000~km/s
centred on the reference redshift listed in
Table~\ref{tab:source}. The uncertainty in the published redshifts is
indicated in each panel. These can be quite substantial.  No high
significance detections of the redshifted CO(1--0) line were
achieved. Upper limits to the integrated line-strength can be
calculated by assuming a representative line-width. Detections of CO
(1--0) in lower redshift samples of ULIRGs (e.g. \citealt{solo97},
\citealt{chun09}) suggest that the typical observed linewidth is about
300~km/s. We therefore assumed a more conservative 500~km/s FWHM for
the determination of an appropriate noise level. (Flux and mass limits
become less restrictive the larger the assumed linewidth.) Since we
are searching for the counterpart of a target with known position and
velocity, the probability for a false positive detection is given by
the tail probability of a standard normal distribution under the
assumption of approximately Gaussian noise. A false positive exceeding
2$\sigma$ then has a probability of 0.023. Even for our entire sample
size of 11 targets, the false positive rate at this level of
significance is only 0.25, which is still significantly less than
unity. The 2$\sigma$ upper limits to integrated linestrength were
determined from the measured fluctuation level over the entire
spectrum (including any possible emission signature) after spectral
smoothing to 500~km/s FWHM. The four cases (00397-1312, 03538-6432,
F04565-2615 and 23515-2917) where the actual line integral within the
central 1000~km/s of the spectrum exceeds this 2$\sigma$ limit are
noted as marginal detections in the Table, together with their
1$\sigma$ errors. As noted above, some of the target redshifts have
large uncertainties, particularly for F00320-3307 and F23555-3436. In
both these cases, there may be evidence for CO emission at an offset
velocity that may well be consistent with the current uncertainties in
the systemic velocity.

In the lower right panel of Fig.~\ref{fig:spectra} we present the
spectrum obtained by aligning all of the measured spectra to relative
velocity and forming the simple unweighted average. This yields a high
signal-to-noise detection of the mean CO(1--0) line from our sample, as
noted in Table~\ref{tab:source}.

The corresponding line luminosity is given by, 
\begin{equation}
L_{CO} = \frac{4 \pi D_L^2\nu_0}{c(1+z)} \int S(V)dV
\end{equation}
or
\begin{equation}
L_{CO} = 1.04\cdot10^{-3}\frac{D_L^2\nu_0}{(1+z)} S_V \ \ \ (L_{\sun})
\end{equation}
for the luminosity distance, $D_L$, in Mpc, the
rest frequency, $\nu_0$, in GHz and the integrated linestrength, $S_V$,
in Jy-km/s. Some authors make use of the quantity,
\begin{equation}
L'_{CO} = \frac{c^2}{2 k_B \nu_0^2}\frac{D_L^2}{(1+z)} \int S(V)dV
\end{equation}
or
\begin{equation}
L'_{CO} = 3.26\cdot10^{7}\frac{D_L^2}{\nu_0^2(1+z)} S_V \ \ \ \rmn{(K-km/s\ pc^2)}
\end{equation}
with $D_L, \nu_0$ and $S_V$ as defined above. For the CO(1--0)
transition these measures are related simply by, $L'_{CO}/L_{CO} =
2.05\cdot10^4$ (K-km/s pc$^{2}$)/L$_{\sun}$. The utility of the
$L'_{CO}$ measure is that it permits more direct comparison of
different line transitions since it is formulated as a product of
brightness temperature with surface area \citep{solo97}. Line emission
of the same brightness temperature originating from the same region
will yield the same $L'_{CO}$, independent of transition or observing
frequency.

Calculation of an associated total hydrogen mass requires an assumed
relationship between this quantity and the CO(1--0) line luminosity. A
good compilation of such conversion factors is given in Fig.~10 of
\citet{tacc08}. There is clearly a large scatter in the derived
conversion factors and there may well be underlying dependancies on
many physical factors, including metallicity, gas surface density and
particularly gas excitation temperature. A plausible conversion factor
appropriate for ULIRGs is likely to be about 1 M$_{\sun}$ per (K-km/s
pc$^{2}$) or about $2\cdot10^4$ M$_{\sun}$ per L$_{\sun}$.

\begin{figure}
\resizebox{84mm}{!}{\includegraphics{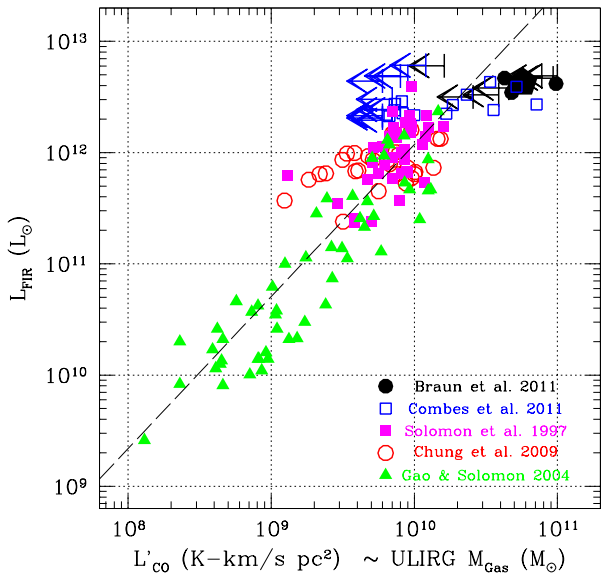}}
\caption{The relationship between CO(1--0) and FIR luminosity. Our
  results are indicated with the filled circles and 2$\sigma$ upper
  limit symbols. The large filled pentagon represents our ``stacked''
  result. A power law of slope 1.37$\pm$0.06 represents a least
  squares fit to the low redshift samples and has a correlation
  coefficient 0.891. }
\label{fig:lpco}
\end{figure}

We plot the distribution of CO(1--0) and FIR luminosity in
Fig.~\ref{fig:lpco}, including both our own results and those of
\citet{comb11} as well as previously published studies of less distant
and less luminous systems by \citet{solo97}, \citet{chun09} and
\citet{gao04}, all of which have been undertaken in the CO(1--0)
transition. Although the low redshift data lie on a well-defined locus
in the plot, the intermediate redshift data from both the current
study and that of \citet{comb11} show divergence from this simple
correlation. A subset of the detections has a significant excess CO
luminosity, while another subset consisting primarily of upper limits
in CO, is significantly underluminous. The difference in CO luminosity
between these two populations is about one order of magnitude. A
power-law with slope of 1.37$\pm$0.06 has been fit by least squares to
the three low redshift samples and is overlaid in the figure. The
correlation coefficient of the data points is 0.891. As noted by
previous authors \citep[eg.][]{chun09}, this relationship is steeper
than linear.  A single power-law slope provides a reasonable
representation of this correlation for FIR luminosities $9.3 <
\log(L_{FIR}) < 12.2$. As noted previously, the associated gas mass for
a given CO(1--0) line luminosity is likely to vary with the gas
kinetic and excitation temperatures, while $L_{FIR}$ is expected to be
very sensitive to the dust temperature, $T_D$, possibly varying as
$T_D^{4-6}$ \citep{soif89}.

\begin{figure}
\resizebox{84mm}{!}{\includegraphics{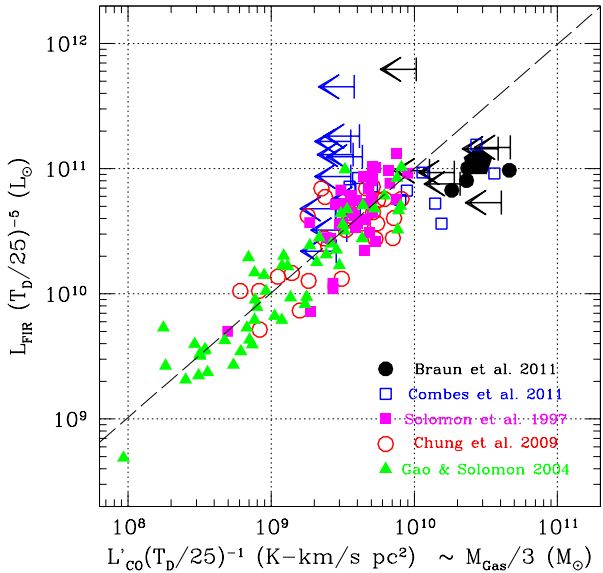}}
\caption{The relationship between temperature normalised CO(1--0) and
  FIR luminosity. $L_{CO}$ is normalised by $(T_D/25)^{1}$ and
  $L_{FIR}$ by $(T_D/25)^{5}$, for a reference dust temperature,
  $T_D$~=~25~K. A power law of
  slope 0.99$\pm0.04$ represents a least squares fit to the low redshift
  samples and has a correlation coefficient 0.902. A similar fit is
  achieved for $L_{CO}$ normalised by $(T_D/25)^{1}$ and
  $L_{FIR}$ by $(T_D/25)^{5.5}$.}
\label{fig:lpcot}
\end{figure}

\citet{gao04} have demonstrated a higher degree of correlation of
$L_{FIR}$ with $L'_{CO}$ once a correction of the FIR luminosity for a
varying dust temperature has been accounted for. The dust temperature
can be estimated from the FIR photometry if an emmissivity law is
assumed. This is often taken to be that of a Planck function
multiplied by $\nu^\beta$, for frequency, $\nu$, and power law
exponent, $\beta$. \citet{lise00} have found that the spectral energy
distributions (SEDs) in their sample of FIR luminous galaxies could be fit
with $0.85~<~\beta~<~1.9$. The dust temperature, in the Wien
approximation, can be estimated from the ratio, $R_S~=~S_1/S_2$, of
flux densities at two fixed observing frequencies, $\nu_1$ and $\nu_2$
from,
\begin{equation}
R_S = R_\nu^{\beta+3} exp\bigg(\frac{h(1+z)(\nu_2-\nu_1)}{kT_D}\bigg)
\end{equation}
where $R_\nu~=~\nu_1/\nu_2$. For the 60 and 100~$\mu$m IRAS bands this yields,
\begin{equation}
T_D = \frac{41.7 (1+z)}{0.22 (\beta + 3) - \log(R_S)} \ \ \  (K)
\label{eqn:td}
\end{equation}
Since the integrated FIR luminosity is expected to scale as
$T_D^{\beta+4}$, we have considered normalisations of the FIR
luminosity down to a reference temperature of 25~K using the form,
\begin{equation}
L'_{FIR} = L_{FIR} \bigg(\frac{T_D}{25}\bigg)^{-(\beta+4)}
\end{equation}
The case has also been made that the CO(1--0) line luminosity would
vary approximately linearly with $T_D$ \citep{solo97} if the dust and
gas temperatures are reasonably coupled, suggesting a similar normalisation of
the form, 
\begin{equation}
L''_{CO} = L'_{CO} \bigg(\frac{T_D}{25}\bigg)^{-\alpha}
\end{equation}
We find empirically,that the values $(\alpha,\beta)$~=~(1,1) minimise
the dispersion in the temperature-normalised CO(1--0) versus FIR
luminosity relationship. The resulting relation is illustrated in
Fig.~\ref{fig:lpcot}, where a power-law of slope 0.99$\pm0.04$ has
been fit by least squares to the three low redshift samples. The
correlation coefficient with this choice of $(\alpha,\beta)$ is
0.902. A very similar relation, with a slope 1.03$\pm0.05$, is
obtained for $(\alpha,\beta)$~=~(1,1.5), which may be more
representative of the FIR SEDs \citep{lise00}. The distinction
between the CO-luminous and CO-poor populations is undiminished in the
temperature corrected relationship.

We can estimate the dust mass associated with our targets from,
\begin{equation}
M_{D} =  \frac{S_{\nu 0} D_L^2}{(1+z)\kappa_{\nu z} B_{\nu z}(T_D)} 
\end{equation}
where $S_{\nu 0}$ is the observed flux density in an FIR band, $D_L$
is the luminosity distance,
$\kappa_{\nu z}$ is the absorption cross section per unit dust mass at
the target rest-frame frequency $\nu_z = \nu_0 (1+z)$ and $B_{\nu
  z}(T_D)$ is the Planck function at this frequency for a dust
temperature, $T_D$ \citep{hild83}. The dust absorption cross section
can be approximated by,
\begin{equation}
\kappa_{\lambda} =  27.2 \bigg(\frac {\lambda_{\mu m}} {100}\bigg)^{-2.15}\big(1+0.625[\log(\lambda_{\mu m})-2.15]^2\big)
\end{equation}
in units of cm$^2$/g for wavelengths expressed in $\mu$m between
40--1000 $\mu$m. This analytic form fits the tabulated data of
\citet{drai03} to better than 5\% over the indicated wavelength
range. We calculate the dust masses for our sample and the comparison
samples noted previously using the observed IRAS 100 $\mu$m fluxes and
the dust temperatures calculated with eqn.~\ref{eqn:td} and
$\beta~=~1.5$, listing these in Table~\ref{tab:source}.

\begin{figure}
\resizebox{84mm}{!}{\includegraphics{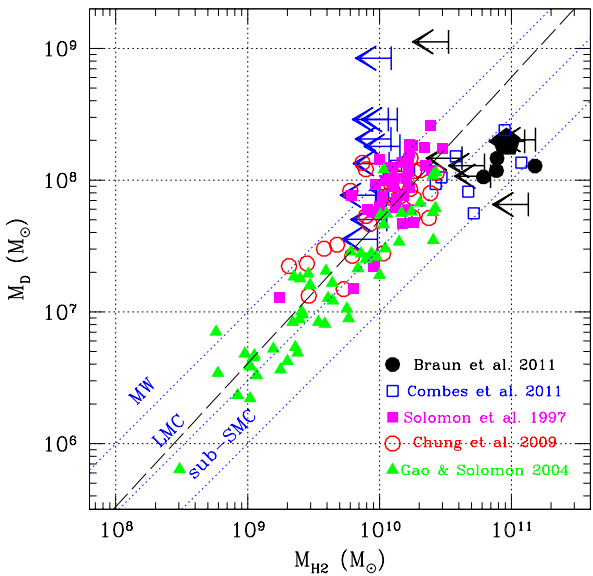}}
\caption{The relationship between molecular hydrogen mass and dust
  mass. A power law of slope
  1.09$\pm0.05$ is shown as a dashed line and represents a least
  squares fit to the low redshift samples and has a correlation
  coefficient 0.893. Dotted lines labelled with sub-SMC, LMC and MW
  represent increasing dust-to-molecular hydrogren mass ratios of
  0.001, 0.003 and 0.01 respectively.}
\label{fig:dvm}
\end{figure}

Once the CO luminosity has been temperature-corrected, it seems more
likely that a single conversion factor to total molecular hydrogen
mass, appropriate to the 25~K reference temperature, might apply. From
\citet{tacc08} this might correspond to about 3 M$_{\sun}$ per (K-km/s
pc$^{2}$) or about $6\cdot10^4$ M$_{\sun}$ per L$_{\sun}$. Even so,
there may well be other systematic dependancies and a large intrinsic
scatter, that add significant uncertainty to this conversion
factor. The relationship between molecular hydrogen mass and dust mass
is shown in Fig.~\ref{fig:dvm}. A power-law fit to the low redshift
samples has slope 1.09$\pm0.05$ and a correlation coefficient of 0.893
and is plotted in the Figure. Also shown are diagonal lines
corresponding to fixed dust to hydrogen mass ratios of 0.001, 0.003
and 0.01, which are labelled sub-SMC, LMC and MW respectively
\citep[cf.][]{drai07}. The sub-SMC designation is used for the lowest
curve since the Small Magellanic Cloud is estimated to have a mass
ratio of 0.002 \citep[eg.][]{wein01}. The correlated locus in Fig.~\ref{fig:dvm} defined by
the low redshift samples systematically increases in dust-to-gas mass
ratio with increasing gas mass, from SMC-like values at the low end,
through Large Magellanic Cloud and up to Milky Way values for $M_{H_2}
> 10^{10}$ M$_\odot$.

What is striking in Fig.~\ref{fig:dvm} is that all of the intermediate
redshift detections in both the current and \citet{comb11} studies
depart from the low redshift trend in the direction of extremely low
dust-to-molecular gas mass ratio. Our well-defined sample average point of
0.0021$\pm$0.0002 is comparable to the dust-to-hydrogen mass ratio of
the SMC. In contrast, there are several upper limits in the figure
which appear to be dust-rich (or gas-poor) by about an order of
magnitude. The most extreme point from the current sample is the
source IRAS 0738-2342, while in the \citet{comb11} sample it is IRAS
19104+8436. Two addtional sources in the Northern sample, [HB89]
1821+643 and F~00415-0737, are also discrepant but less extreme. While
very little is currently known about F~00415-0737, \citet{ruiz10} have
recently presented SED fits for IRAS 0738-2342 and [HB89]
1821+643. They estimate relative AGN:Starburst contributions to the
bolometric luminosity of about 50:50 and 80:20 for these two sources. Both
IRAS 19104+8436 and [HB89] 1821+643 are classified as Seyfert
1/QSOs, suggesting that $L_{FIR}$ for IRAS 19104+8436 may also be
dominated by non-thermal rather than dust emission. 

It seems plausible that the apparent ``gas poor'' sub-population seen
at intermediate redshift in Figs.~\ref{fig:lpco}, ~\ref{fig:lpcot} and
~\ref{fig:dvm} may simply be a manifestation of AGN contamination,
with the star formation dominated population demonstrating a
well-defined trend relative to the low redshift samples, consistent
with a dramatic decline in the dust-to-molecular gass mass ratio.

\begin{figure}
\resizebox{84mm}{!}{\includegraphics{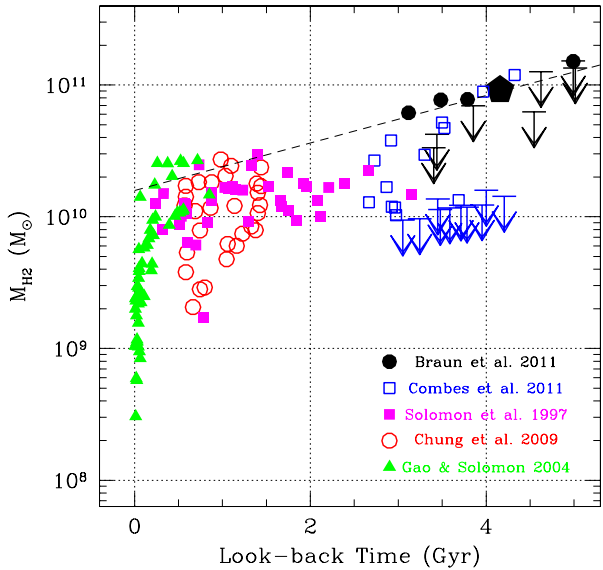}}
\caption{The relationship between look-back time and ULIRG molecular
  hydrogen mass. An indicative
  upper envelope to molecular gas mass, $\log(M_{H_2}) = 10.2 + 0.18
  \tau$, is plotted demonstrating the order of magnitude decline in
  maximum ULIRG gas mass over the past 5 Gyr. }
\label{fig:mvt}
\end{figure}

We consider the distribution of ULIRG molecular hydrogen mass with
look-back time, $\tau$, in Fig.~\ref{fig:mvt}. Since ULIRGs are
identified by their extreme luminosity in the FIR band, where
self-opacity and foreground extinction effects are minimal, they
represent a population that is easily recognised in unbiased surveys
of the sky. Both our own study, as well as the others shown in the
Figure, were drawn from the most luminous soures in the IRAS all-sky
survey. As such they are not expected to suffer from incompleteness at
high luminosity and despite the fact that they are a rare population, they
permit a useful assessment of the associated molecular gas mass that
is required to feed the ULIRG phenomenon. Although the lowest detected
gas masses at each look-back time simply reflect the sensitivity of
the observation and the cutoff $L_{FIR}$ of the sample, the upper
envelope to mass in Fig.~\ref{fig:mvt} is a significant attribute of
the population, at least once the sample extends over a representative
volume of about 10$^{7}$ Mpc$^3$; since the most luminous sources are
clearly the easiest to detect at any redshift. This condition should
be met for redshifts greater than about 0.03, or look-back times
exceeding 0.4 Gyr. This local volume saturation effect is apparent in
the Figure for $\tau < 0.5$ Gyr. The indicative curve drawn in the Figure,
\begin{equation}
\log(M_{H_2}) =  10.2 + 0.18 \tau 
\end{equation}
demonstrates the order of magnitude decline in maximum molecular gas
mass of the ULIRG population during the past 5 Gyr. Such a decline is
comparable to, but even more dramatic, than that seen in the star
formation rate density \citep{hopk06} which can be well described by,
\begin{equation}
\log(\dot \rho_*) = -1.76\pm0.02 + 0.131\pm0.004 \tau
\end{equation}
for a star formation rate density $\dot \rho_*$ in
M$_\odot$yr$^{-1}$Mpc$^{-3}$ for $\tau <$ 8 Gyr. This contrast suggests
significant changes in the molecular gas mass function
at even these modest look-back times.

We will present further analysis of the time evolution of molecular
gas and other baryonic constituents of galaxies in a subsequent publication.

\section*{Acknowledgments}

We thank an anonymous referee for their constructive suggestions of
improvements to the manuscript, including formation of a ``stacked''
spectrum for the sample. The Mopra radio telescope is part of the
Australia Telescope National Facility which is funded by the
Commonwealth of Australia for operation as a National Facility managed
by CSIRO.

\label{lastpage}

\end{document}